\documentstyle[12pt,aaspp4]{article}

\def\approx{$\sim$}
\def\persqcm{$\rm cm^{-2}$}
\def\error{$\pm$}
\def\e#1{$\times 10^{#1}$}
\def\tenup#1{10$^{#1}$}
\def\asec{\arcsec}
\def\amin{\arcmin}
\def\kms{km~s$^{-1}$}
\def\solmass{$\rm M_{\sun}$}

\begin{document}

\title{Improved Searches for HI in Three Dwarf Spheroidal Galaxies}
\author{L. M. Young}
\affil{Astronomy Department, New Mexico State University,
Box 30001, Las Cruces, NM 88003; lyoung@nmsu.edu}

\begin{abstract}   
Previous searches for HI in our Galaxy's dwarf spheroidal companions
have not been complete enough to settle the question of whether or not
these galaxies have HI, especially in their outer parts.
We present VLA observations of three dwarf spheroidals:  Fornax, Leo~II,
and Draco, all of which have known stellar velocities.
The new data show no HI emission or absorption.
Column density limits in emission are 4\e{18}--7\e{18}~\persqcm\ in the
centers of the galaxies.
The importance of the new observations is that they
cover larger areas than previous searches and they
are less plagued by confusion with foreground (Galactic) HI.
The apparent absence of neutral gas in the Fornax dwarf spheroidal is
especially puzzling because recent photometry shows evidence of stars only
\tenup{8} years old.
We discuss whether the VLA observations could have missed significant
amounts of HI.

\end{abstract}


\section{Introduction}

Dwarf spheroidal galaxies, the smallest companions of our own Milky Way
galaxy, were long thought to be old and dead galaxies devoid of any 
interstellar medium.
They show no sign of current star formation.
Furthermore, searches for HI emission from the dwarf spheroidal galaxies (Knapp et al.
1978; Mould et al. 1990; Koribalski, Johnston, \& Otrupcek 1994)
found no evidence of gas in the galaxies,
with one possible exception (Carignan et al. 1998).
Optical and UV absorption experiments for Leo~I (Bowen et al. 1995, 1997)
also detected no gas.
Thus, it is commonly assumed that the dwarf spheroidal galaxies have no
neutral ISM at all.

However, recent work on color-magnitude diagrams of the Local Group dwarf 
spheroidals contradicts this picture of old, dead galaxies.
Most of the dwarf spheroidals experienced periods of star
formation activity at various times from 10~Gyr to 1~Gyr ago (e.g. 
Smecker-Hane 1997 and references therein; Hurley-Keller, Mateo, \& Nemec
1998).
The Fornax dwarf spheroidal even seems to contain some very young stars, 
\approx \tenup{8} yrs old (Stetson, Hesser, \& Smecker-Hane 1998).
Since star formation requires neutral gas, these facts clearly demonstrate that 
the spheroidals have had an interstellar medium (ISM) in the past several Gyr
or less.
This ISM could have existed in the spheroidals for most of their lifetimes,
or it could have been captured from some external source such as the
Magellanic Stream, high velocity clouds, or a cooling intracluster medium
(e.g. Silk, Wyse, \& Shields 1987).  Regardless of its origin,
star formation histories show that the spheroidals had significant amounts of neutral gas in the recent past.
Therefore, they might be expected to have an interstellar medium today.

The question of whether there is neutral gas in the dwarf spheroidals 
has important implications for 
our understanding of star formation and the evolution of galaxies.
If the dwarf spheroidals had neutral gas in the past but they have none now, 
what happened to the gas?
Two popular ideas are that
(1) the neutral gas in the spheroidals may have been stripped by
interactions with the outer halo of our Galaxy, or that
(2) a burst of star formation activity, with attendant supernovae and 
stellar winds, may have evacuated the neutral gas from the galaxies.  
Skillman \& Bender (1995) discuss some advantages and disadvantages of
these ideas.
Another possibility might be that if the gas density were 
lowered, neutral gas could be ionized by the interstellar UV field
(e.g. Bland-Hawthorn, Freeman, \& Quinn 1997; Bland-Hawthorn 1998).
In any case it is clear that the presence or absence of neutral gas in the
dwarf spheroidals is an important clue to their history.

The existing data on the HI content of dwarf spheroidals is not complete
enough to answer the question of whether they have any neutral gas
(Section~\ref{why}).
Therefore, we have conducted new searches for HI in and around
the Leo~II, Fornax, and Draco 
dwarf spheroidal galaxies using the VLA's D configuration.
These observations cover larger regions than before and limit possible
confusion with foreground Galactic HI.
Subsequent sections describe what was previously known about the HI
contents of dwarf spheroidals, the current observations, and the
implications of these new data.

\section{Existing Data on the HI Content of Dwarf Spheroidals}\label{why}

It is commonly repeated that dwarf spheroidals have no HI, but
an examination of the published results shows that the data are
simply not adequate to make this conclusion.
The basic problem is incompleteness.
HI surveys (Hartmann 1996; Huchtmeier \& Richter 1986) have covered a large
fraction of the Northern sky or have searched many galaxies, but at 
poor sensitivity.
Searches for optical and UV absorption lines in front of quasars near
Leo~I (Bowen et al. 1995, 1996) provide extremely low column density
limits,
but they only probe three points at radii of 3, 5, and 10 times
the tidal radius of the galaxy. 
(In this paper, core and tidal radii for the dwarf spheroidals are 
taken from the work of
Irwin \& Hatzidimitriou [1995].)
Even the published HI observations (Knapp et al. 1978; Mould et al. 1990;
Koribalski et al. 1994) are inconclusive, for reasons described below.

Existing searches for HI in dwarf spheroidal galaxies suffer from two problems.
The major problem with existing single-dish observations 
is that they searched only a small fraction of the galaxies'
areas.
In the case of the Draco and Ursa Minor spheroidals (Knapp et
al. 1978) a beam with a half-power radius of 5\arcmin\ was centered 
on galaxies whose core radii (semi-major axes) are 9\arcmin\ and 16\arcmin.
Therefore, less than one third of the area inside the core radii of these
galaxies has been observed.
Similar arguments apply to the Sagittarius dwarf spheroidal, Fornax, and
Carina (Koribalski et al. 1994; Knapp et al. 1978; Mould et al. 1990).
The small HI mass limits given for these galaxies are commonly
misunderstood and misused
because they apply only to the small area which has been observed, not to
the entire galaxy.
Neutral gas could be present in the unobserved parts of the optical galaxies.

Gas could also be present in the outer parts beyond the optical galaxies.
Blow-out models (e.g. Dekel \& Silk 1986; De Young \& Heckman 1994; Mac~Low \& Ferrara
1997)
provide some reasons why gas might be found in the outer parts
of quiescent dwarf galaxies instead of in the center.
Furthermore, several dwarf galaxies are indeed observed to have
HI minima centered on the galaxy and
HI rings (or partial rings) outside the optical galaxy. 
These include M81~dwarf~A (Sargent, Sancisi, \& Lo
1983; Puche \& Westpfahl 1994), Sag~DIG (Young \& Lo 1997),
and even the dwarf spheroidal Sculptor (Carignan et al. 1998).
It is not clear whether the blow-out models mentioned above 
explain the observed HI rings. 
In any case 
it cannot be assumed that all of the gas should be in the centers 
of the dwarfs, 
where previous searches have been made.

Another problem with single-dish observations of the dwarf spheroidals 
is that HI gas at velocities close to 0 \kms\ could
have been overlooked because of confusion with Galactic HI.
The best example of how this can happen is the case of the Phoenix dwarf
(sometimes referred to as a ``transition" galaxy between irregulars and
spheroidals).
Phoenix was observed with a single-dish telescope (Carignan et al. 1991)
and subsequently with the VLA (Young \& Lo 1997).
The VLA observations detected a cloud of HI at $-23$ \kms,
but the single-dish observations did not detect the cloud 
because this velocity lies
under partially-subtracted Galactic HI which the VLA resolves away.
More recent interferometric observations of Phoenix show even more HI than
Young \& Lo (1997) found (Carignan 1998, private communication).
Knapp's (1978) observations of Fornax and Leo~II might also 
suffer from confusion with Galactic HI.
The problem is exacerbated by the fact that the velocity of Leo~II was not
known at the time the observations were made.
Leo~II's optical velocity is +76~\error~1 \kms\ (Vogt et al. 1995), but at that position the
Galactic HI extends out to velocities of almost +120 \kms\ (Young \&
Gallagher 1998).
Thus, the detection efforts made to date are limited and cannot
give a conclusive answer about 
whether the spheroidals really have no HI.

\section{Observations}

We address some of the problems of previous observations by using the NRAO 
Very Large Array (VLA)\footnote{The National Radio Astronomy Observatory
is a facility of the National Science Foundation, operated under
cooperative agreement by Associated Universities, Inc.}
to search for HI emission in and
around the Fornax, Leo~II, and Draco dwarf spheroidal galaxies.
For Draco, these VLA observations cover a much larger fraction of the
galaxy than has been previously searched. 
For Leo~II and Fornax, these VLA observations suffer much less from
confusing local HI emission.
One reason that there is less confusion in the VLA images 
is that most of the local HI emission is simply not detected.
Unlike a single-dish telescope, the VLA acts as a high-pass spatial filter. 
Most of the radiated
power from foreground Galactic HI is on relatively low spatial
frequencies (large angular scales of degrees or greater).
Thus, most of the Galactic HI does not appear in the VLA images.
In addidition, spatial mapping allows us to distinguish gas that is probably not
associated with the galaxy.

The observational setups are described in Table~\ref{obstable}.
The observations were made in the D and DnC configurations in 1997--1998.
They cover a bandwidth of 1.56 MHz, which gives a usable velocity range of
about 290 \kms\ centered close to the
optical velocity of the galaxy as determined from stellar absorption
lines.
The velocity resolutions were 2.6 \kms, based on previous experience
detecting HI clouds in the vicinity of the Phoenix and Tucana dwarf
spheroidals (Young \& Lo 1997; Oosterloo et al. 1996).
The primary beam of the VLA at 21cm has a full width at half maximum of
31\amin, i.e. a response of 50\% at a radius of 15.5\amin, and a response
of 10\% at a radius of 26.4\amin\ (Napier \& Rots 1982).
The data were mapped using natural weight and again with a tapering weight
function which emphasizes large spatial structures, both
before and after continuum subtraction.
Continuum emission was subtracted directly from the combined dataset 
using the task UVLIN in the AIPS package.
Table~\ref{obstable} gives the positions, velocity ranges covered, 
beam sizes,
noise levels, and column density limits for these observations.
The beam linear sizes are computed assuming the distances given in Irwin
\& Hatzidimitriou (1995).

The phase/pointing centers of the VLA observations, given in
Table~\ref{obstable}, are quite close to the actual centers of the
galaxies.
The phase/pointing centers for Leo~II and Draco are less than 1\arcmin\ away
from the galaxy centers given in Irwin \& Hatzidimitriou (1995).
The Fornax dwarf spheroidal is observed to have significant asymmetrical
structure, with the peak stellar density about 6\arcmin\ northeast of the
centroid of the lowest isophotes (Stetson et al. 1998).
The VLA phase/pointing center is between the peak stellar density and the
galaxy centroid, about 2\arcmin\ southwest of the position of peak stellar density.
The center velocities in Table~\ref{obstable} are also within 6 \kms\ of the most
recently determined heliocentric stellar velocities, which are 53~\error~2 \kms\ for
Fornax (Mateo et al. 1991), 76~\error~1 \kms\ for Leo~II (Vogt et al. 1995),
and $-$294~\error~3 \kms\ for Draco (Hargreaves et al. 1996).

\section{Results}

We find no evidence for any HI emission or absorption that can be
associated with the dwarf spheroidal galaxies.
Sensitivity limits for HI emission are given in Table~\ref{obstable}
and are typically 5\e{18}~\persqcm\ at the galaxy center,
twice that at the VLA half-power point (radius 15.5\arcmin), and
5\e{19}~\persqcm\ at the VLA 10\% power point (radius 26.4\arcmin).
The column density limits are given as the column density of a 3$\sigma$
signal in three consecutive channels.
For purposes of comparison, the low column density HI clouds observed 
near the Sculptor, Phoenix, and Tucana dwarfs peak at
2\e{19}~\persqcm\ (Carignan et al. 1998), 4\e{19}~\persqcm\ (Young \& Lo
1997), and 8\e{19}~\persqcm\ (Oosterloo et al. 1996), respectively.
We argue in Section~\ref{discussion} that 
there is not likely to be a significant amount of HI which we have not
detected, especially in the centers of the galaxies.

\begin{figure}
\plotone{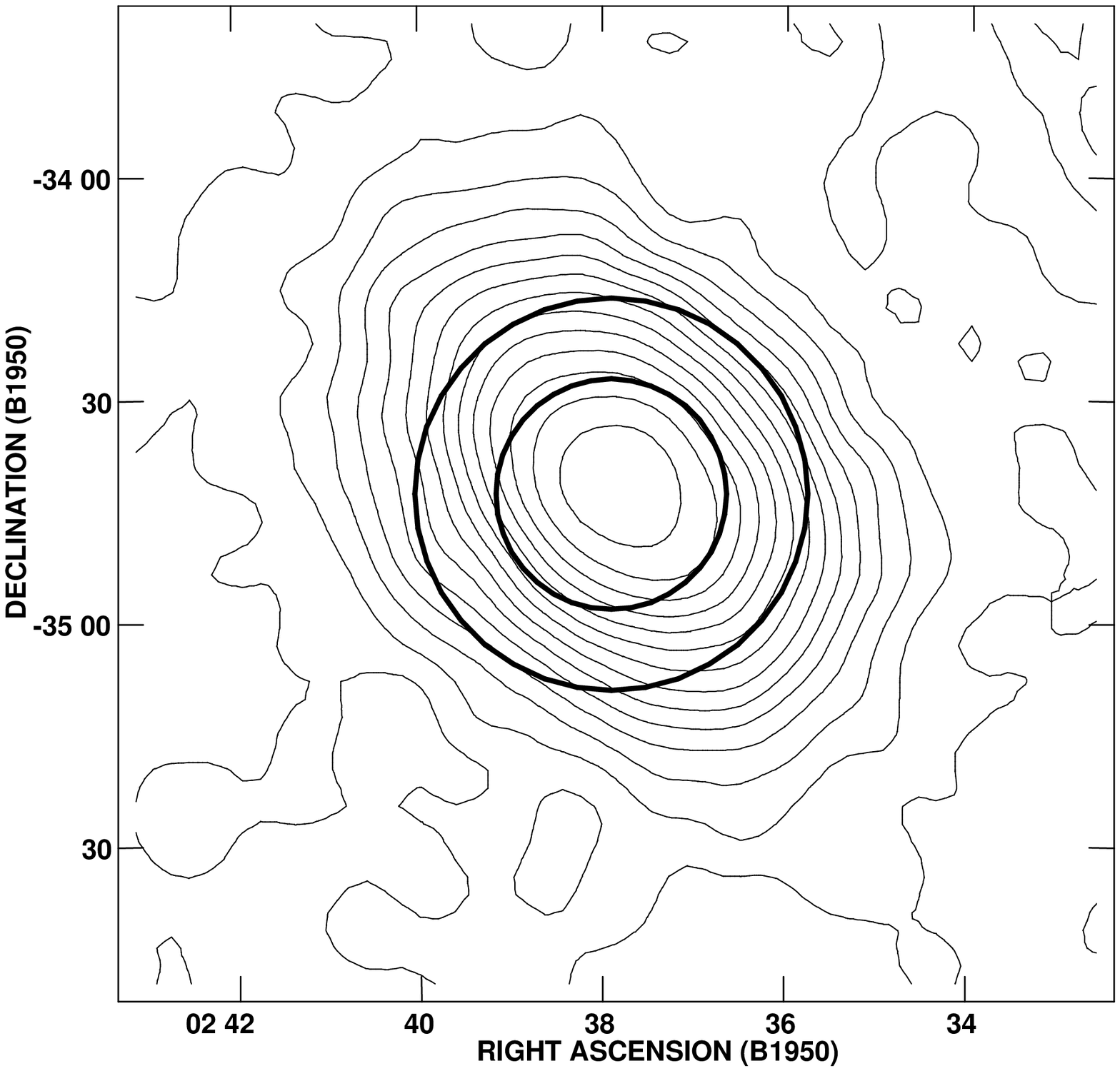}
\caption{Optical extent of the Fornax dwarf spheroidal galaxy compared to
the VLA field of view.
The light contours show the isopleth map of Irwin \& Hatzidimitriou (1995),
with the same contour levels described in that paper.
The heavy circles show the VLA half-power points (radius 15.5\arcmin) and
10\% power points (radius 26.4\arcmin).
\label{fornaxfig}
}
\end{figure}

\begin{figure}
\plotone{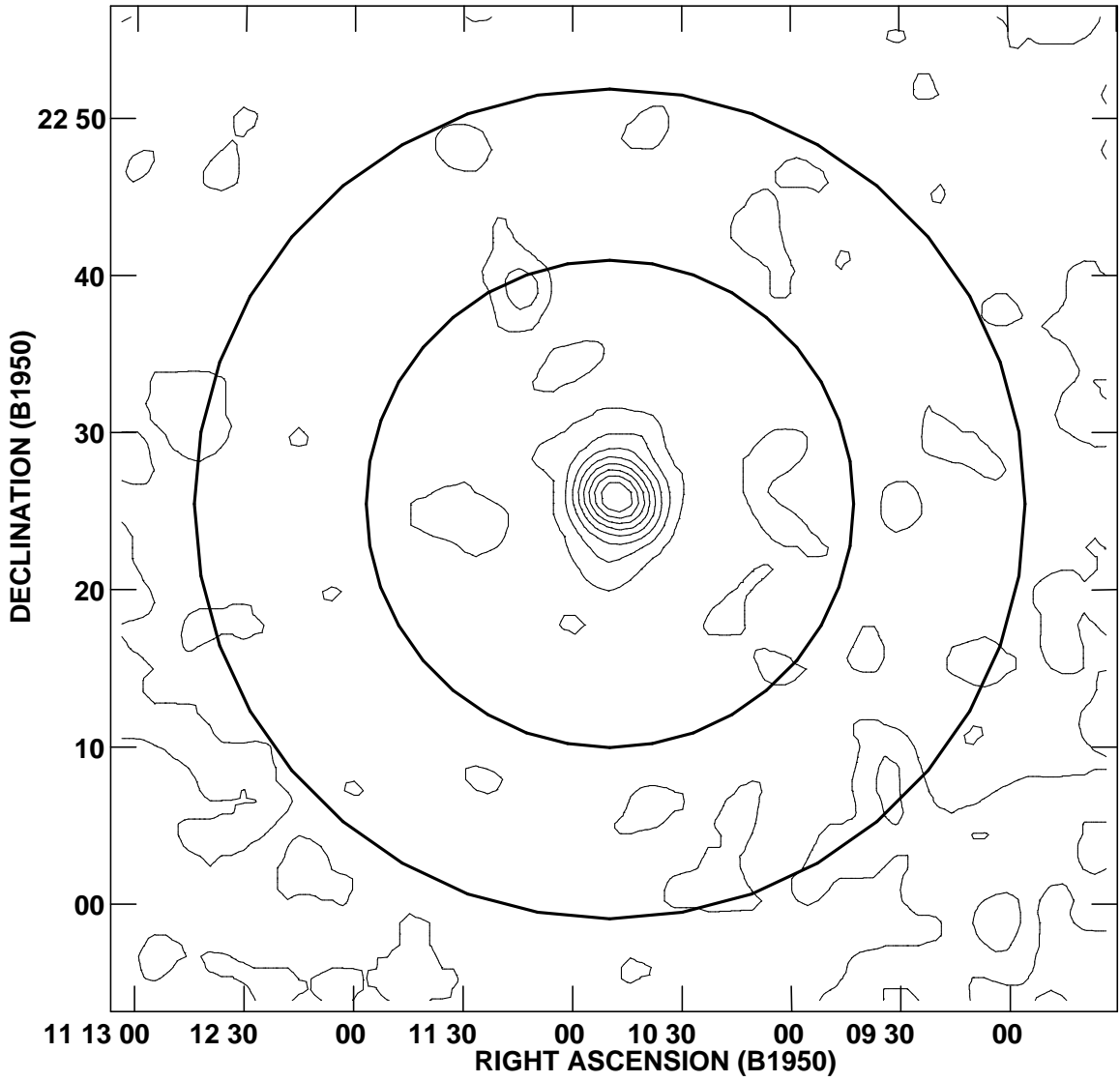}
\caption{Same as Figure~\ref{fornaxfig}, for Leo~II.
\label{leo2fig}
}
\end{figure}

\begin{figure}
\plotone{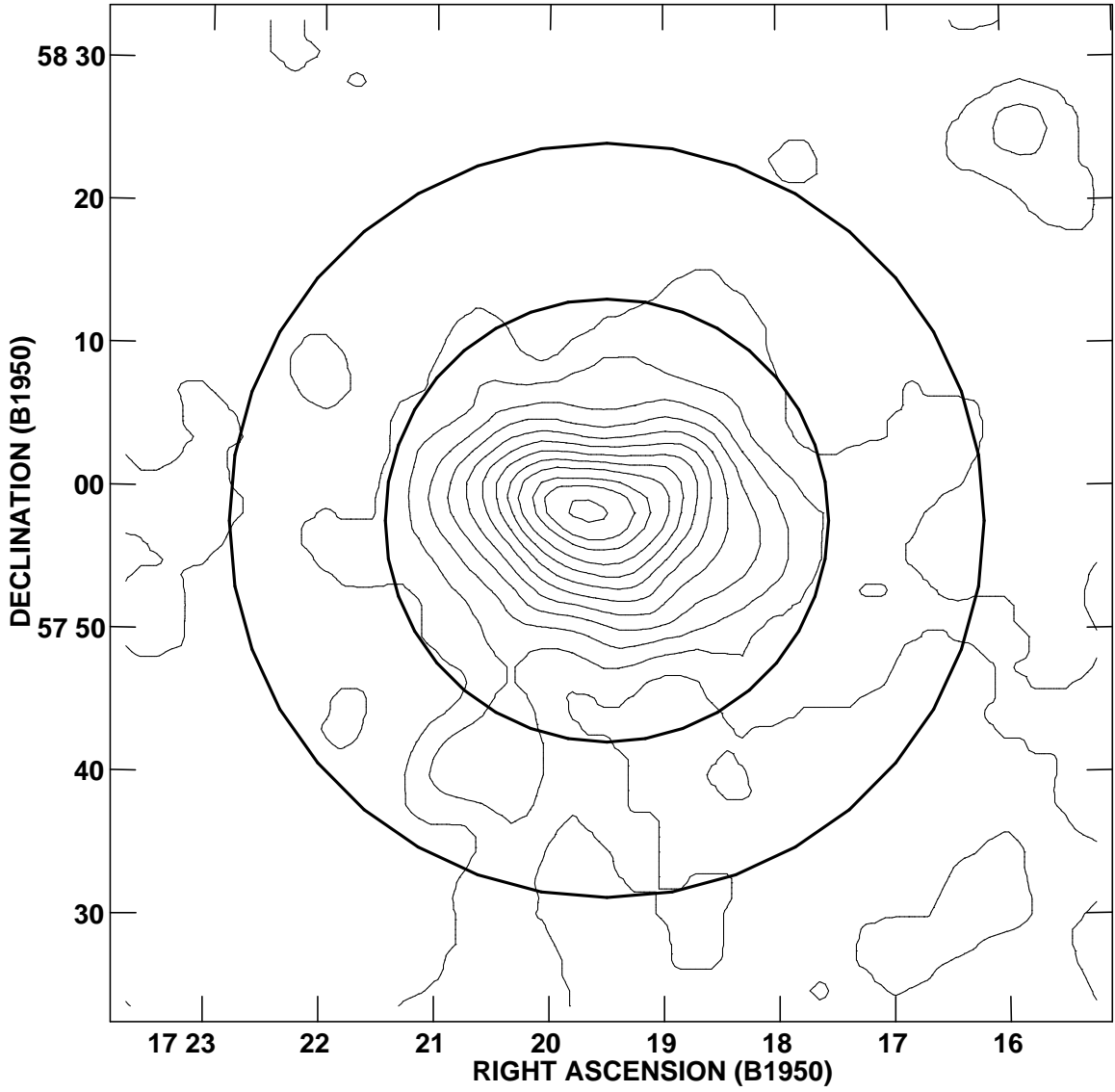}
\caption{Same as Figure~\ref{fornaxfig}, for Draco.
\label{dracofig}
}
\end{figure}

Figures \ref{fornaxfig}, \ref{leo2fig}, and \ref{dracofig} show isopleth
maps of the galaxies (Irwin \& Hatzidimitriou 1995) along with the
half-power and 10\% power circles of the VLA primary beam.
Table~\ref{obstable} gives the major axis core radii and tidal radii of these
galaxies, taken from Irwin \& Hatzidimitriou (1995).
For Leo~II, the column density limit is 4.3\e{18}~\persqcm\ at the center
of the galaxy and 5.2\e{18}~\persqcm\ at the tidal radius of the galaxy.
Thus, effectively all of Leo~II has been searched at good sensitivity
(see also section \ref{discussion}).
For Draco, the detection limit is 7.1\e{18}~\persqcm\ at the center of
the galaxy, increasing to 8.7\e{18}~\persqcm\ at the (major axis) core
radius, and about 7\e{19}~\persqcm\ at the tidal radius.
Since the tidal radius of the galaxy is approximately equal to the VLA's
10\% power radius,
any HI associated with the galaxy would most likely
be within the VLA field of view. 
However, the sensitivity at Draco's tidal radius is probably not good
enough to exclude the presence of HI there (section \ref{discussion}).
In the case of Fornax, the detection limit is 4.6\e{18}~\persqcm\ at the
galaxy center and 7.9\e{18}~\persqcm\ at the core radius.
The galaxy's tidal radius (71\amin) is much larger
than even the VLA's 10\%-power radius.
Thus, there might still be undetected HI somewhere between the core
radius and the tidal radius of Fornax or Draco.

\begin{figure}
\plotone{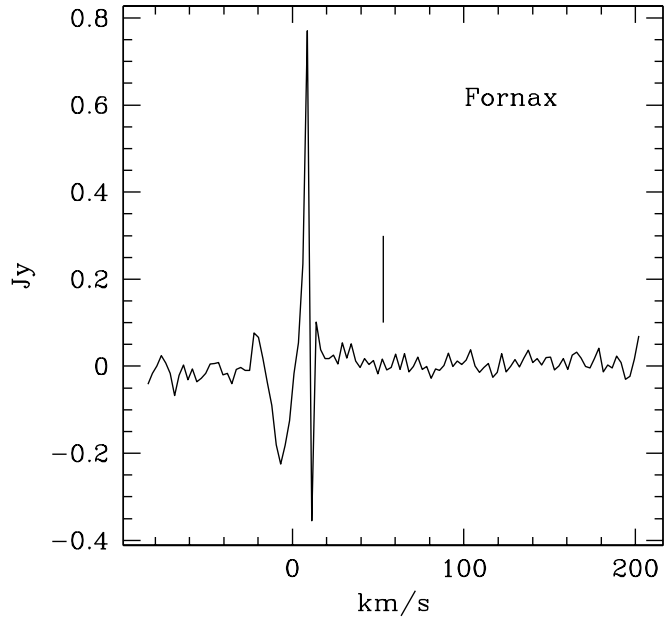}
\caption{Spectrum from Fornax's data cube.
This spectrum is a sum over the area inside the half-power point, after
correction for the primary beam.
The optical velocity of the galaxy is marked with a vertical line.
The strong positive and negative features near 0 \kms\ (heliocentric)
arise from foreground Galactic HI.
\label{fornaxspect}
}
\end{figure}

\begin{figure}
\plotone{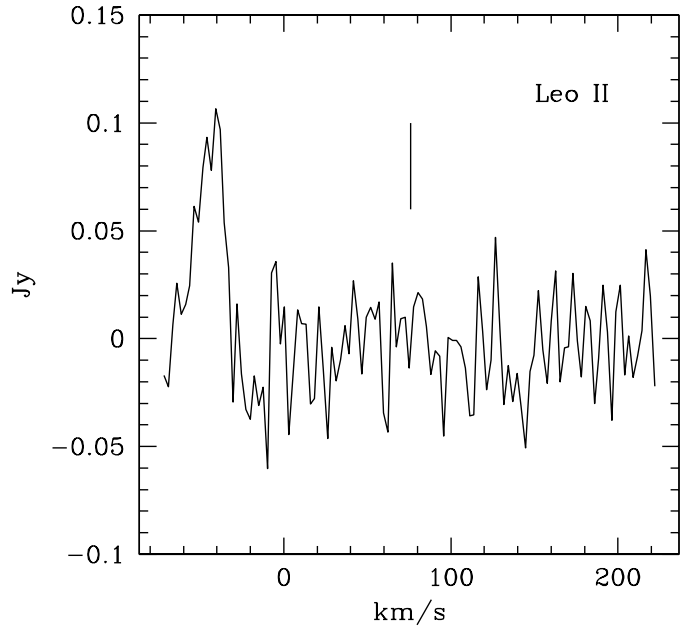}
\caption{Similar to Figure~\ref{fornaxspect}, for Leo~II.
The features at velocities lower than 0 \kms\ are caused by
Galactic HI.
\label{leo2spect}
}
\end{figure}

\begin{figure}
\plotone{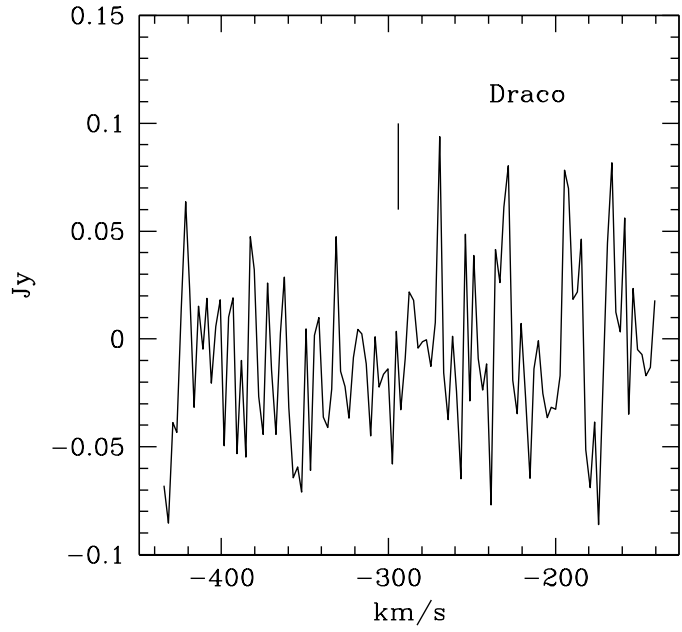}
\caption{Similar to Figure~\ref{fornaxspect}, for Draco.
\label{dracospect}
}
\end{figure}

We find some HI emission in the data cubes, but it is undoubtedly from
foreground Galactic HI.
Galactic emission in these VLA images takes the form of large-scale
(\approx 20\arcmin) positive and negative features at velocities near 0
\kms; the negative
features arise because the VLA has resolved out most of the total flux.
Figures \ref{fornaxspect}, \ref{leo2spect}, and \ref{dracospect} show
spectra constructed from each data cube and illustrate that the emission
features are not associated with the dwarf galaxies in velocity.
In the case of the Fornax dwarf the Galactic HI is seen at velocities between 19
and $-$22 \kms\ (heliocentric), with greatest intensities at two 
peaks at about 9 and $-$4 \kms; the galaxy's stellar velocity is +53\error 2 \kms\
(Mateo et al. 1991).
Near Leo~II, Galactic HI is observed between velocities of 3 and $-$59
\kms, with greatest intensities at $-$2 and $-$41 \kms, in agreement 
with a single-dish spectrum of Leo~II (Young \& Gallagher 1998).
The stellar velocity of Leo~II is +76\error 1 \kms\ (Vogt et al. 1995).
No Galactic HI is observed in the Draco field.
The VLA has effectively removed contamination from Galactic HI at the
velocities of the dwarf galaxies, so we conclude that we have not missed
any dwarf galaxy emission which is hiding behind Galactic HI.

There are a number of background continuum sources in each of the VLA
fields, but no absorption is detected towards any of them.
Table~\ref{abs-limits} presents the positions, peak flux densities, and
optical depth limits (3$\sigma$) for the brightest continuum 
sources in each field.
The positions quoted are simply those of the brightest pixels
and should not be assumed to be more accurate than
0.5 pixel (5\asec--10\asec).
The last column also gives the distance between the continuum source and the
galaxy center. 
Since none of the continuum sources are particularly bright, the column
density limits in absorption are not as meaningful as the limits in
emission.
For example, the smallest optical depth limit is 0.11 for a source
22\arcmin\ away from the center of Draco; for this source, the column
density upper limit is $\rm N_{HI} < 5.2\times 10^{17}\,T_S\ cm^{-2}$,
and a typical spin temperature $\rm T_S$ of 100~K would give column density
limits of 5\e{19}~\persqcm.

\section{Discussion}\label{discussion}

Interferometers, by their nature, cannot detect very smooth spatial structures.
However, it is unlikely that HI in the dwarf spheroidal galaxies has
escaped detection by reason of being too smooth for the VLA to detect.
The present datasets include baselines as short as 170~$\lambda$
(20\arcmin) for all three galaxies, and the VLA can be expected to
image structures as large as 15\amin\ (Perley 1997).
At the adopted distances of these spheroidals, 15\amin\ corresponds to 
linear sizes of 520 pc (Fornax), 900 pc (Leo~II), and 310 pc (Draco).
Thus, HI in the spheroidals could be ``resolved out" by the VLA only if
it was very smooth on scales smaller than at least 300--900 pc.
Such a situation would be highly unusual, as every other galaxy which has been
observed in HI emission at high resolution shows small scale structures.
Kalberla et al. (1985) found structures as small as 0.5 pc -- 1 pc in
Galactic HI; a mosaic of HI in the Small Magellanic Cloud shows intricate
structure down to scales of 30 pc (Staveley-Smith et al. 1997).
If HI structures are due in large part to star formation
activity, the dwarf spheroidals might indeed have relatively smooth 
interstellar media.
However, the scales involved, at least 300 pc, are so large that 
the absence of any structure seems a remote possibility.
Furthermore, while smoothly distributed gas could not be detected in
emission, it {\it could} be detected in absorption against the point
sources, and no absorption was found.

We consider it possible, but unlikely, that HI could exist in the dwarf
spheroidals at column density levels below the sensitivities achieved in
these VLA obervations, especially in the galaxy centers.
HI simply does not seem to exist at low column density
levels in the outer parts of galactic systems.
Sensitive observations of a spiral galaxy (van Gorkom 1993) show that the HI
disk of the spiral cuts off sharply when the HI column density reaches
about \tenup{19}~\persqcm.
A similar effect is seen in high velocity clouds, where Colgan et al.
(1990) observe a tendency for the HI in the clouds to cut off sharply at
column densities below 5\e{18}~\persqcm.
Corbelli \& Salpeter (1993a, 1993b) and others have argued that these HI cutoffs are
probably caused by ionization by the galactic and/or extragalactic UV radiation field.
In this picture, hydrogen could not exist in neutral form in the dwarf 
spheroidals at column densities below about \tenup{19}~\persqcm.
(And because of the high spatial resolution of these VLA images, 30--75 pc, 
the column density of any HI should not be diluted much by a
small beam filling factor.) 
The observed column density limit for Leo~II is well below
\tenup{19}~\persqcm\ even beyond the tidal radius of the galaxy;
therefore, it is unlikely that Leo~II contains significant amounts of
HI.
For Draco and Fornax, the column density limits rise above
\tenup{19}~\persqcm\ between the core radius and the tidal radius.
For these galaxies, it is highly unlikely that there is significant HI 
within the core radii, but we cannot rule out the presence of HI at column
density levels of a few times \tenup{19}~\persqcm\ between the core radius
and the tidal radius.

\section{Implications}\label{implications}

\begin{figure}
\plotone{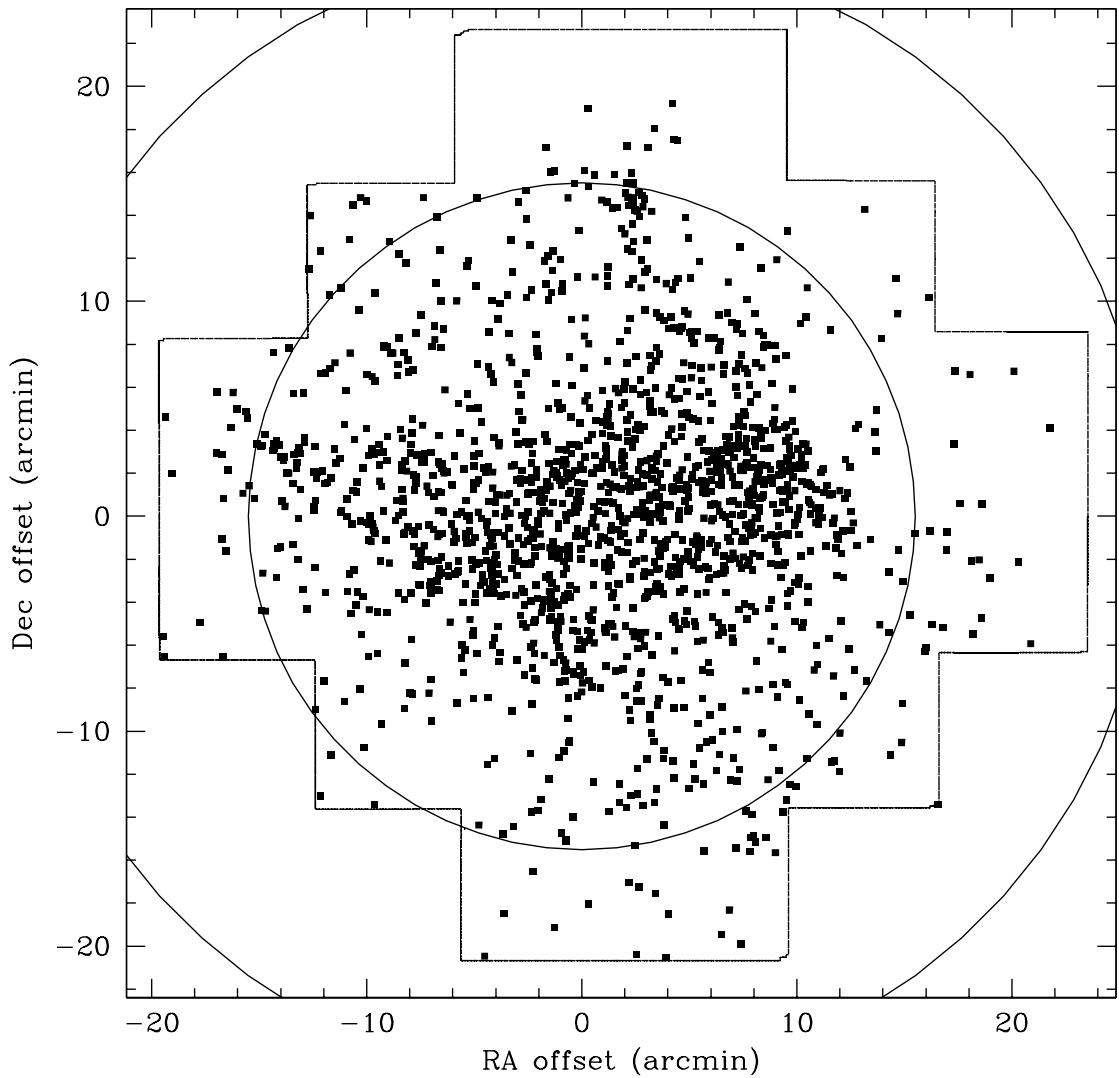}
\caption{Distribution of the bright blue stars in Fornax.
Filled symbols show the locations of stars which, in the data of Stetson et
al. (1998), have magnitudes $18.5 < (B+R)/2 < 22.0$ and colors 
$B-R < 0.25$.
(These criteria are likely to include some blue 
evolved stars in addition to massive main sequence stars.)
The two circles show the VLA half-power points and 10\% power points,
and the cross-shaped region is the field observed by Stetson et al.
The axis units are arcminutes with respect to the VLA pointing center.
North is up and east is to the left, as in Figure~\ref{fornaxfig}.
\label{bluestars}
}
\end{figure}

The stellar population of Leo~II is predominantly made up of stars with
ages between 7 and 14 Gyr, and
the stars in Draco are at least 10 Gyr old (Smecker-Hane 1997 and
references therein; Grillmair et al. 1998).
However, recent observations of Fornax (Stetson, Hesser, \& Smecker-Hane
1998) indicate that there are a number of young stars in that galaxy,
so that the absence of neutral gas in the center of Fornax
becomes an interesting puzzle.
The photometry of Stetson et al. (1998) reveals a large number of bright blue
($B-R<0$) stars which are interpreted as a young main sequence with an
age of only 100 to 200 million years.
These young stars are concentrated in the center of the galaxy, with
a distribution much like that of the bulk of the stars in Fornax.
Figure~\ref{bluestars}, which is based on the data of Stetson et al.
(1998), shows the distribution of the bright blue stars in Fornax and in the
VLA field of view.

The young stars in Fornax are concentrated in the center of the VLA field of view;
and as little as \tenup{8} years ago, these stars must have been associated
with neutral gas.
We infer two possibilities:  either (1) the gas that formed the young stars
in Fornax
is now ionized or molecular, and has not been detected; and/or
(2) the neutral gas and the young stars parted company in the last
\tenup{8} years.
Perhaps the neutral gas was ejected from the galaxy, as in the popular
``blow-out" models (e.g. Mac~Low \& Ferrara 1998, and references therein).
Since the one-dimensional velocity dispersion of the stars in Fornax is 
11\error 2 \kms\ (Mateo et
al. 1993), gas moving at the escape speed of 38 \kms\ would reach the
outer edge of the VLA field of view (26\amin\ = 920 pc) after only 2.4\e{7}
yr.
Apparently, enough time has elapsed to get rid of the gas which formed the
young stars.
If neutral hydrogen existed at some point in the past,
and then expanded because of blow-out caused by star formation, 
its column density could drop significantly and
it might now exist in an ionized state at very low emission measure.
The presence of these young stars and the apparent absence of neutral gas
is a puzzle which we cannot resolve at this time.

\section{Summary}

We present VLA searches for HI in the Fornax, Leo~II, and Draco dwarf
spheroidal galaxies.
No HI was detected in these galaxies, either in emission or absorption.
In all three cases the VLA observations cover larger areas than have been
previously searched, and for Fornax and Leo~II the new data have the
important advantage of removing possible confusion with Galactic HI.
For Leo~II, the column density limit in emission is 5\e{18}~\persqcm\ out
to the tidal radius.
For Fornax and Draco the column density limits are 4\e{18} and
7\e{18}~\persqcm\ in the galaxy centers, increasing to
\tenup{19}~\persqcm\ at points between the core radii and the tidal
radii.
In the Draco dwarf galaxy we also find HI optical depth limits $\tau<0.1$
towards two continuum sources at 1.9 and 2.4 core radii from the center.
From these observations we conclude that there is no significant HI within
the tidal radius of Leo~II or in the centers of Fornax and Draco.
It will be necessary to observe still larger areas to determine whether
there is HI in the outer parts of Fornax and Draco.
However, these observations are much more complete than previous ones,
and they close important loopholes in assessing the question of
whether there is or isn't HI in the spheroidals.

\acknowledgments

Thanks to J. Gallagher for helpful discussions, to M. Irwin for
providing the isopleth maps of the dwarf spheroidals, and to P. Stetson for
providing data on the bright blue stars in Fornax.


\newpage

\newpage

\begin{deluxetable}{llll}
\tablecaption{VLA Observations of HI in Dwarf Spheroidals\label{obstable}}
\tablewidth{0pt}
\tablehead{
\colhead{} & \colhead{Fornax} & \colhead{Leo II} & \colhead{Draco}
}
\startdata
phase center (J2000) &  02 39 58.5  &  11 13 29.2  &  17 20 18.7  \nl
             &  $-$34 30 00   &  +22 09 11   &  +57 54 48   \nl
total time, hr &  9.0  &  5.0  &  2.0 \nl
configuration &  DnC  &  D  &  DnC  \nl
date &  15oct97, 19oct97  &  08jan98  &  23oct97 \nl
phase calibrator & J0240$-$231  &  J1120+143  &  J1634+627  \nl
bandpass calibrator & J0137+331  &  J1331+305  &  J0137+331 \nl
center velocity (\kms) & +55.0  &  +70.0  & $-$290.0 \nl
beam size (\asec)& 121$\times$73 & 75$\times$71 & 95$\times$67 \nl
beam linear size (pc)& 70$\times$42 & 75$\times$71 & 33$\times$23 \nl
rms noise (mJy/beam) & 1.58  &  0.89  &  1.74  \nl
rms noise (K)        &  0.11 &  0.10  &  0.17  \nl
N$_{\rm HI}$ limit (\persqcm) & 4.6\e{18} & 4.3\e{18} & 7.1\e{18} \nl
N$_{\rm HI}$ limit (\solmass\,pc$^{-2}$)  & 0.037 & 0.035 & 0.057 \nl
distance (kpc) & 120\error 8  &  207\error 10  &  72\error 3  \nl
core radius (\arcmin)  &  14  &  2.9  &  9.0 \nl
tidal radius (\arcmin) &  71  &  8.7  &  28 \nl
\enddata
\end{deluxetable}

\begin{deluxetable}{rrrrcrr}
\tablecaption{Optical Depth Upper Limits for Continuum Sources\label{abs-limits}}
\tablewidth{0pt}
\tablehead{
\colhead{Galaxy} & No. & \colhead{RA} & \colhead{Dec} & \colhead{Peak} & 
\colhead{$\tau$} & \colhead{Distance} \nl
\colhead{} & \colhead{} & \multicolumn{2}{c}{J2000.0} & \colhead{mJy/beam} & 
\colhead{} & \colhead{\arcmin} \nl
}
\startdata
Draco & 1 & 17 20 58.8 & +57 49 08 & 32 & 0.16 & 8.0 \nl
      & 2 & 17 19 04.3 & +58 04 27 & 18 & 0.29 & 13.6 \nl
      & 3 & 17 20 51.3 & +57 55 18 & 44 & 0.12 & 17.3 \nl
      & 4 & 17 20 49.8 & +57 35 38 & 19 & 0.28 & 19.9 \nl
      & 5 & 17 18 21.1 & +58 10 05 & 49 & 0.11 & 21.6 \nl
\nl
Leo\,II & 1 & 11 13 59.4 & +21 59 11 & 8.7 & 0.30 & 12.9 \nl
        & 2 & 11 14 12.3 & +21 57 26 & 14 & 0.19 & 16.1 \nl
        & 3 & 11 12 07.1 & +22 10 40 & 15 & 0.18 & 18.5 \nl
        & 4 & 11 12 15.9 & +22 00 40 & 8.7 & 0.30 & 18.8 \nl
\nl
Fornax & 1 & 02 40 08.2 & $-$34 29 20 & 12 & 0.38 & 1.0 \nl
       & 2 & 02 39 03.4 & $-$34 36 39 & 15 & 0.31 & 14.7 \nl
\enddata
\end{deluxetable}


\begin{thebibliography}{}

\bibitem[]{} Bland-Hawthorn, J. 1998, in Galactic Halos: A UC Santa Cruz
Workshop, ed. D. Zaritsky (San Francisco: ASP), 113

\bibitem[]{} Bland-Hawthorn, J., Freeman, K. C., \& Quinn, P. J.  1997,
\apj, 490, 143

\bibitem[]{} Bowen, D. V., Blades, J. C., \& Pettini, M.  1995, \apj, 448,
634
 
\bibitem[]{} Bowen, D. V., Tolstoy, E., Ferrara, A., Blades, J. C., \&
Brinks, E.  1997, \apj, 478, 530

\bibitem[]{} Carignan, C., Demers, S., \& C\^ot\'e, S.  1991, \apj, 381,
L13 

\bibitem[]{} Carignan, C., Beaulieu, S., C\^ot\'e, S., Demers, S., \&
Mateo, M.  1998, \aj, in press

\bibitem[]{} Colgan, S. W. J., Salpeter, E. E., \& Terzian, Y.  1990, \apj,
351, 503

\bibitem[]{} Corbelli, E., \& Salpeter, E. E.  1993a, \apj, 419, 94

\bibitem[]{} Corbelli, E., \& Salpeter, E. E.  1993b, \apj, 419, 104

\bibitem[]{} De Young, D. S., \& Heckman, T. M.  1994, \apj, 431, 598

\bibitem[]{} Dekel, A., \& Silk, J.  1986, \apj, 303, 39
 

\bibitem[]{} Grillmair, C. J., Mould, J. R., Holtzman, J. A., Worthey, G.,
Ballester, G. E., et al. 1998, \aj, 115, 144

\bibitem[]{} Hargreaves, J. C., Gilmore, G., Irwin, M., J., \& Carter, D.
1996, \mnras, 282, 305

\bibitem[]{} Hartmann, D., \& Burton, W. B. 1997, Atlas of Galactic Neutral
Hydrogen (Cambridge: Cambridge U. P.)

\bibitem[]{} Huchtmeier, W. K., \& Richter, O. G.  1986, \aaps, 1986, 63

\bibitem[]{} Hurley-Keller, D., Mateo, M., \& Nemec, J.  1998, \aj, 115,
1840

\bibitem[]{} Irwin, M. J., \& Hatzidimitriou, D.  1995, \mnras, 277, 1354

\bibitem[]{} Kalberla, P. M. W., Schwarz, U. J., 
\& Goss, W. M.  1985, \aap, 144, 27

\bibitem[]{} Koribalski, B., Johnston, S., \& Otrupcek, R.  1994, \mnras, 270, 43

\bibitem[]{} Knapp, G. R., Kerr, F. J., \& Bowers, P. F.  1978, \aj, 83, 360

\bibitem[]{} Mac Low, M.-M., \& Ferrara, A. 1998, \apj, in press

\bibitem[]{} Mateo, M., Olszewski, E., Welch, D. L., Fischer, P., \&
Kunkel, W.  1991, \aj, 102, 914

\bibitem[]{} Mateo, M., Olszewski, E. W., Pryor, C., Welch, D. L., \& Fischer, P.
1993, \aj, 105, 510

\bibitem[]{} Mould, J. R., Bothun, G. D., Hall, P. J., Staveley-Smith, L.,
\& Wright, A.  E.  1990, \apj, 362, L55

\bibitem[]{} Napier, P. J., \& Rots, A. H.  1982, Very Large Array Test
Memorandum No. 134

\bibitem[]{} Oosterloo, T., Da Costa, G. S., \& Staveley-Smith, L.  1996,
\aj, 112, 1969

\bibitem[]{} Perley, R. A.  1997, Very Large Array Observational Status Summary

\bibitem[Puche \& Westpfahl 1994]{puc94} Puche, D., \& Westpfahl, D.
1994, in {Proceedings of the ESO/OHP
Workshop on Dwarf Galaxies}, eds. G. Meylan and P. Prugniel (Garching: ESO),
273

\bibitem[]{} Sargent, W. L. W., Sancisi, R., \& Lo, K. Y.  

\bibitem[]{} Silk, J., Wyse, R. F. G., \& Shields, G. A.  1987, \apj, 322,
L59

\bibitem[]{} Skillman, E. D., \& Bender, R. 1995, RevMexAA (Serie de
Conferencias), 3, 25

\bibitem[]{} Smecker-Hane, T. 1997, in ``Star Formation--- Near and Far" eds. S. Holt and
L. G. Mundy (AIP: New York), 571.


\bibitem[]{} Staveley-Smith, L., Sault, R. J., Hatzidimitriou, D.,
Kesteven, M. J., \& McConnell, D.  1997, \mnras, 289, 225

\bibitem[]{} Stetson, P. B., Hesser, J. E., \& Smecker-Hane, T.  1998, \pasp, 110, 533

\bibitem[]{} van Gorkom, J. 1993, in The Environment and Evolution of
Galaxies, eds J. M. Shull \& H. A. Thronson Jr (Dordrecht: Kluwer), p. 345

\bibitem[]{} Vogt, S. S., Mateo, M., Olszewski, E. W., \& Keane, M. J.
1995, \aj, 109, 151

\bibitem[]{} Young, L. M., \& Gallagher, J. S. III 1998, in preparation

\bibitem[]{} Young, L. M., \& Lo, K. Y.  1997, \apj, 476, 127 

\end{thebibliography}
\end{document}